\documentstyle[preprint,aps]{revtex}
\tighten
\draft
\widetext
\preprint{CLNS 98/1553, HUTP-98/A056, NUB 3183}
\bigskip
\bigskip
\begin{document}
\title{Brane World}
\medskip
\author{Zurab Kakushadze$^{1,2}$\footnote{E-mail:
zurab@string.harvard.edu} and S.-H. Henry Tye$^{3}$\footnote{E-mail:
tye@mail.lns.cornell.edu}}
\bigskip
\address{$^1$Lyman Laboratory of Physics, Harvard University, Cambridge,
MA 02138\\
$^2$Department of Physics, Northeastern University, Boston, MA 02115\\
$^3$Newman Laboratory of Nuclear Studies, Cornell University, Ithaca, NY 14853}
\date{September 15, 1998}
\bigskip
\medskip
\maketitle

\begin{abstract}

{}In string theory, the brane world scenario, where the Standard Model gauge 
and matter fields live inside some branes while gravitons live in the bulk, 
can be a viable description of our universe. In this note we argue that the 
brane world actually is a likely description of nature. Our discussion
includes a revisit of certain issues, namely, coupling unification, 
dilaton stabilization and supersymmetry breaking, in the context of the 
brane world scenario. In particular, we discuss various possible string 
scenarios and their phenomenological implications in the brane world framework.

\end{abstract}

\section{Introduction}

{}The observed universe contains gauge and matter fields as well as gravity. 
Since string theory\footnote{In the following we will mostly discuss critical string theory whose target space-time is ten dimensional. Certain strong coupling regimes of string theory can be described by eleven dimensional M-theory.} 
is the only known theory that incorporates consistent quantum gravity, one 
would like to see how string theory can describe our universe. 
In string theory gauge fields (as well as the corresponding charged matter) 
and gravity can be incorporated in rather distinct ways. 
Thus, gravity lives in ten dimensional space-time, whereas
gauge fields can be localized on $p\leq 9$ spatial dimensional extended 
objects known as $p$-branes. In particular, one can imagine a scenario where our four dimensional world (including the strong and electroweak interactions as well as quarks and leptons) resides inside 
a $p$-brane (or a set of overlapping branes) with extra $p-3\geq 0$ spatial dimensions compactified on a manifold with some finite volume $V_{p-3}$. However, four dimensional gravity is free to propagate in the ten dimensional bulk of the space-time with the remaining $9-p$ spatial dimensions compactified on a manifold with some finite volume $V_{9-p}$. For appropriate
values of $V_{p-3}$ and $V_{9-p}$ all the known experimental constraints appear to be satisfied, and this scenario, which we will refer to as ``Brane World'', {\em a priori} seems to be a viable possibility for describing our universe.     

{}In fact, the brane world picture appears to be favored by various string theoretic considerations in the context of phenomenology. More precisely, for string theory to describe our universe generically (that is, without appealing to fine tuning or any other possible ``special circumstances'') it seems that our four dimensional world must reside inside a $p$-brane (or intersection thereof with other branes). Moreover, the corresponding string theory is likely to be strongly coupled (or, more precisely, the string coupling $g_s
{\ \lower-1.2pt\vbox{\hbox{\rlap{$>$}\lower5pt\vbox{\hbox{$\sim$}}}}\ } 1$, and the theory
appears not to have a dual weak coupling description). These appear to be generic phenomenological predictions of string theory.
In this note we attempt to give certain evidence in support of these statements. Our discussions are mostly based on the following observations which we revisit in the present context:\\
({\em i}) the gauge and gravitational couplings must unify at the string scale;\\ 
({\em ii}) dilaton, whose VEV determines the string coupling, must be stabilized, which cannot be achieved in perturbative string theory;\\ 
({\em iii}) the low energy effective field theory describing the Standard Model model is weakly coupled at the electroweak scale.\\
Thus, in the brane world picture, the four dimensional gauge and 
gravitational couplings 
scale as $1/V_{p-3}$ and $1/V_{p-3} V_{9-p}$, respectively. By tuning $V_{9-p}$ (for $p<9$) we can achieve gauge and gravitational coupling unification. Furthermore, dilaton stabilization must occur via some non-perturbative dynamics which generically implies $g_s{\ \lower-1.2pt\vbox{\hbox{\rlap{$>$}\lower5pt\vbox{\hbox{$\sim$}}}}\ } 1$. Then for $p>3$, we can choose $V_{p-3}$ large enough so that the four dimensional gauge couplings are small even if 
$g_s$ is not. 
The $3<p<9$ feature means that our world is inside a $p$-brane.  
In this brane world picture, the string scale $M_s$ can {\em a priori} be 
anywhere between the electroweak scale and the Planck scale.
Here we would like to stress that the brane world is a {\em simplified} (semi-classical) picture (if the corresponding string theory is indeed strongly coupled), so it should not be taken as a precise description of nature. Rather, it appears to be an adequate starting point which might help orient our thinking
about how string theory might describe our universe. As to practical applications, we may hope that various dualities provide tools to understand at least some qualitative (and, perhaps, even some quantitative) aspects of such a brane world.

{}Our analyses here are intended to be as model independent as possible. In particular, we will discuss various possible scenarios in which non-supersymmetric low energy effective field theory can emerge. This is especially relevant for the discussion of dilaton stabilization.
We also point out that, in the context of non-perturbative string theory, 
the dilaton VEV (and so the string coupling $g_s$) may vary point-wise in the 
compact directions. This might have interesting phenomenological consequences.

{}Let us summarize some other issues discussed in this paper. 
Thus, we point out that the dilaton cannot be stabilized 
perturbatively even if supersymmetry is broken at the 
tree level. Here we consider the following scenarios 
for supersymmetry breaking. First, supersymmetry can 
be broken at the tree level both on the brane and in the 
bulk. Here we consider tachyon free as well as tachyonic 
cases. Second, we can have ``Brane Supersymmetry'' 
where we have tree-level supersymmetry on the brane 
but the bulk is non-supersymmetric. Finally, supersymmetry 
breaking might occur dynamically on the brane, and then 
can be transmitted to the bulk.  Dilaton stabilization must be 
due to some non-perturbative (gauge) dynamics on (some) branes, 
and this generically requires the corresponding string vacuum to be 
strongly coupled. This raises the question of why is the effective field 
theory corresponding to the Standard Model weakly coupled. 
We point out that in the brane world picture this can be achieved by 
simply having the Standard Model fields and the gauge theory 
responsible for dilaton stabilization arise from different sets of branes. 
In contrast, the above problem still remains an open question in, say, 
the old perturbative heterotic framework.  

{}The remainder of this note is organized as follows. In section II we discuss the constraints imposed by gauge and gravitational coupling unification. These considerations already indicate that generically our four dimensional world should reside inside a $p$-brane.
In section III we discuss various possible scenarios where the resulting low energy spectrum of string theory is non-supersymmetric as required by experiment. This is intended
to facilitate the discussion in section IV where we consider various scenarios of dilaton stabilization, which lead us to conclude that the string coupling generically is expected to be $g_s {\ \lower-1.2pt\vbox{\hbox{\rlap{$>$}\lower5pt\vbox{\hbox{$\sim$}}}}\ } 1$. To obtain a theory with a weakly coupled sector in the low energy effective field theory, it seems
necessary to have the brane world picture. We also discuss the possibility of varying dilaton VEV, and its plausible phenomenological implications. F-theory provides an example of such a scenario. In section V we give our conclusions. In particular, we give a qualitative description of how a brane world scenario may describe our universe.

\section{Gauge and Gravitational Coupling Unification}

{}In this section we review the discussion in \cite{witten} which leads to the conclusion that it is easier to satisfy the constraint of gauge and gravitational coupling unification in the brane world picture than in the perturbative heterotic framework. 
In fact, the coupling unification\footnote{Such a unification does not 
necessarily require that gauge couplings for all gauge groups
are identical at the string scale and equal the properly defined 
dimensionless gravitational coupling. Rather, generically one can allow 
these couplings to be different by numerical coefficients of order one such 
as the current algebra levels.}
indicates that {\em generically} the brane world picture seems to be a 
natural framework in this context.

{}To be specific, let us carry out this discussion in the context of Type I (or Type I$^\prime$) string theory where we have some number of parallel D$p$-branes $p\leq 9$ on which open strings can end \cite{polchin}. Thus, gravity lives in the ten dimensional bulk, and the gauge theory lives in the world-volume of the D$p$-branes\footnote{The brane world picture in the effective field theory context was discussed in \cite{early}.}. 
The four dimensional Planck scale and the four dimensional gauge coupling (at the string scale) are given by\footnote{The following are tree level relations which receive quantum corrections such as threshold corrections. We will ignore such corrections at the moment as generically they do not change the following qualitative picture.}    
\begin{eqnarray}
 &&M_P^2=G_N^{-1}={8 \over (2\pi)^{6} g_s^{2}} M_s^8 V_{p-3} V_{9-p}~,\\
 &&{1\over g_{\small{gauge}}^{2}}={1\over (2\pi)^{p-2} g_s} M_s^{p-3} V_{p-3}~.
\end{eqnarray}
Here $G_N$ is the four dimensional Newton's constant, $M_s=1/\sqrt{\alpha^\prime}$ is the string scale. For later convenience we can parametrize the volumes as follows\footnote{If $p=3$ then $V_{p-3}=1$, and similarly $V_{9-p}=1$ if $p=9$.} : $V_{p-3}=(2\pi r/M_s)^{p-3}$, and $V_{9-p}=(2\pi {\widetilde r}/M_s)^{9-p}$. Here dimensionless $r$ and ${\widetilde r}$ measure (in the units of the string length $l_s=\sqrt{\alpha^\prime}=M_s^{-1}$) the sizes of the compact dimensions inside and transverse to the $p$-branes, respectively. (Note, however, that these quantities do not literally correspond to the ``compactification radii'' as the compact six dimensional manifold need not be toroidal.) Without loss of generality, we can assume\footnote{Here
we should point out that by $V_{p-3}$ and $V_{9-p}$ we mean effective compactification volumes {\em after} taking into account world-sheet corrections. Thus, if the corresponding sigma model is strongly coupled (so that we cannot compute these effective volumes), we should go to a dual picture where the $\alpha^\prime$ corrections can be reliably computed
(or else it is not even clear what one means by ``gauge coupling unification''). 
In certain cases (such as toroidal orbifolds) this can be achieved via T-duality.} 
that $r,{\widetilde r}\geq 1$. 

{}Let us first consider the case of 9-branes, where we have the following relations:
\begin{eqnarray}\label{D9}
 && M_s={\alpha} M_P r^3/\sqrt{2}\geq {\alpha} M_P/\sqrt{2}~,\\
 && g_s =4 M_s^2/ M_P^2\alpha~,
\end{eqnarray}  
where $\alpha =g^2_{\small{gauge}}/4\pi$ is the ``grand unification'' coupling constant. In the case of the Standard Model gauge group coming from 9-branes we do not expect (just as in perturbative heterotic superstring \cite{kap}) large threshold corrections to the above formulas in the context of perturbative Type I string theory. In particular, the unified gauge coupling is expected to be $\alpha\approx 1/24$, and the unification scale should be $M_{\small{GUT}}\approx 2\times 10^{16}~{\mbox{GeV}}$ \cite{gut}\footnote{These values correspond to gauge coupling unification in the minimal supersymmetric Standard Model (MSSM) with the assumption of minimal matter content up to $M_{\small{GUT}}$ and superpartner thresholds at $\sim 1 ~{\mbox{TeV}}$.}. In fact, we can assume that $M_s\sim M_{\small{GUT}}$. However, this is not 
compatible with (\ref{D9}) which predicts $M_s{\ \lower-1.2pt\vbox{\hbox{\rlap{$>$}\lower5pt\vbox{\hbox{$\sim$}}}}\ } 3\times 10^{17}{\mbox{GeV}}$ \cite{CKM}. To lower the string scale we would have to assume that $r<1$. (In particular, to have $M_s\approx M_{\small {GUT}}$ we must take $r\sim .4$.) However, we can then T-dualize and describe the resulting theory in the context of Type I$^\prime$ with D8-branes. 

{}If $p<9$, then we have two parameters $r$ and ${\widetilde r}$ which can 
be tuned to obtain $M_s\approx M_{\small {GUT}}$. Moreover, the string 
coupling can be either large\footnote{For large string coupling some caution is required in using the following relations, since they can receive large quantum corrections.} or small\footnote{In this note, we will sloppily refer to  
$g_s {\ \lower-1.2pt\vbox{\hbox{\rlap{$>$}\lower5pt\vbox{\hbox{$\sim$}}}}\ } 1$
as the strong coupling regime. The precise definition of the loop expansion 
parameter is $N\lambda_s$ for the open string sector,
and $\lambda_s^2$ for the closed string sector, where $N$ is the number of the
corresponding D-branes, and $\lambda_s = g_s/4\pi$. Thus, a world-sheet with 
$b$ boundaries (corresponding to D-branes), $g$ handles (corresponding to 
closed string loops) and $c$ cross-caps (corresponding to orientifold planes)
is weighted with $(N\lambda_s)^b \lambda_s^{2g-2+c}$.} as it is controlled by 
$r$, whereas $M_s$ depends on both $r$ and ${\widetilde r}$ (where $M_P$ and 
$\alpha$ are assumed to be fixed) 
\begin{eqnarray}\label{Dp}
 && M_s=\alpha M_P (r^{p-3}/{\widetilde r}^{9-p})^{1/2}/\sqrt{2}~,\\
 && g_s =2\alpha r^{p-3}~.
\end{eqnarray} 
In fact, by a suitable choice of $r$ and ${\widetilde r}$ we can {\em a priori} bring the string scale below $M_{\small{GUT}}$. In the limiting case where
$M_s \sim 1~{\mbox{TeV}}$, we have the TeV scale scenario \cite{lyk,TeV}.   

{}Here the following remark is in order. If $M_s<M_{\small {GUT}}$ we immediately face the problem of gauge coupling unification. If we assume that MSSM is not a correct description above some intermediate scale (between $M_{\small {GUT}}$ and $\sim 1~{\mbox{TeV}}$) we can have unification which occurs at a scale different than $M_{\small {GUT}}$. Thus, some additional states charged under the Standard Model gauge group may have such intermediate masses, and above the corresponding scale the gauge coupling running would be modified resulting in a different value for $M_{\small {GUT}}$ (as well as for $\alpha$) than given above. Generically such effects raise the value of $M_{\small {GUT}}$ \cite{DF}\footnote{Here one could attempt to use such a scenario to ameliorate the discrepancy between $M_s$ and $M_{\small{GUT}}$ in the case of D9-branes. However, the situation here is similar to that in perturbative heterotic superstring, and to obtain gauge and gravitational coupling unification one generically has to
 appeal to fine tuning 
(such as judicious choice of additional matter fields and their masses).}. However, if sizes (which we will sloppily keep on calling $r$) of some of the compact directions within the world-volume of D$p$-branes are large compared with $1/M_s$, the evolution of gauge couplings above the Kaluza-Klein threshold $1/r$ is no longer logarithmic but power-like \cite{dien}. (This can alternatively be thought of as considering higher dimensional gauge theory at energy scales larger than $1/r$.) This observation was used in \cite{dien} to argue that gauge coupling unification might be able to occur at the scale (which would be identified with the string scale) much lower than $\sim10^{16}~{\mbox{GeV}}$. (Note that this would ultimately require $p>3$.) In fact, {\em a priori}, this scale can be anywhere between 
$\sim10^{16}~{\mbox{GeV}}$ and $\sim1~{\mbox{TeV}}$. However, unification compatible with precision experimental data (for such quantities as $\sin^2\theta_W$) requires a judicious choice of quantum numbers (with respect to the Standard Model gauge group) of the Kaluza-Klein modes \cite{dien}. At present no explicit string vacuum with such properties is known, and it remains an open question (which would ultimately have to be addressed in any scenario with $M_s{\ \lower-1.2pt\vbox{\hbox{\rlap{$<$}\lower5pt\vbox{\hbox{$\sim$}}}}\ }10^{16}~{\mbox{GeV}}$) whether the gauge coupling unification constraint can be satisfied\footnote{Here we should mention that there may be an alternative to solving the gauge coupling unification problem \cite{ST}. Thus, if some of the Standard Model (or of an extension thereof) gauge symmetry comes from a different set of branes, then the gauge couplings (at the string scale) need not be the same for all gauge groups. Generically, however, some 
degree of fine tuning would be necessary 
in such a scenario as the gauge couplings on different branes depend on different compactification volume factors.}. (For other recent works on TeV scale string/gravity
scenarios, see, {\em e.g.}, \cite{related}. 
TeV scale compactifications were studied in \cite{quiros} in
the context of supersymmetry breaking.)    

{}Finally, we would like to compare brane world string unification with the old perturbative heterotic superstring picture. In the latter case the gauge and gravitational interactions both come from the closed string sector so that both $G_N$ and $\alpha$ are proportional to $g^2_s$. (Recall that, say, in the case of Type I, gravity comes from the closed string sector so that $G_N\sim g_s^2$, whereas gauge interactions come from the open string sector so that $\alpha\sim g_s$.) Moreover, we have only one volume factor $V_6$ corresponding to the internal six dimensional compactification space. This results in the relation $M_s =\sqrt{\alpha} M_P/2$ which predicts the value for $M_s$ to be a factor of $10-20$ (the uncertainty is due to possible threshold effects) higher than $M_{\small{GUT}}$. This is the essence of the unification problem in the old heterotic framework\footnote{{\em A priori} the unification problem 
can be ameliorated
in the perturbative heterotic framework if one considers grand unification within the effective field theory. In such a scenario the Standard Model gauge couplings unify at $M_{\small{GUT}}\approx
2\times 10^{16}~{\mbox{GeV}}$, and above this scale we have a {\em simple} grand unified gauge group whose coupling unifies with the dimensionless gravitational coupling at  $M_s$
which is somewhat higher than $M_{\small{GUT}}$. However, the price one has to pay for implementing such a scenario is that generically certain degree of fine tuning is required to solve other problems (such as proton stability) associated with grand unification \cite{three}.}. 
The brane world approach has the advantage that it is less constraining. 

{}On the other hand, we can ask whether in strongly coupled heterotic string the corresponding prediction is changed. Here we need to distinguish between two cases.
Strongly coupled $Spin(32)/{\bf Z}_2$ heterotic superstring is dual to Type I string theory which we have already discussed. As to the $E_8\otimes E_8$ case, its strong coupling regime corresponds to an M-theory compactification on $(S^1/{\bf Z}_2)\times X_6$, where $S^1$ corresponds to the eleventh dimension of M-theory, and $X_6$ is a six dimensional compact manifold \cite{HW}. In this case we have two parameters to tune: $R$ and ${\widetilde R}$, where $(2\pi R)^6={\mbox{Vol}}(X_6)$, and ${\widetilde R}$ is the size of the eleventh dimension. Gravity now propagates in the eleven dimensional bulk (so that $G_N$ depends on both $R$ and ${\widetilde R}$), whereas the $E_8\otimes E_8$ gauge fields live on two ten dimensional components\footnote{These can be thought of as $9+1$ dimensional walls (reminiscent of 9-branes) embedded in eleven dimensional space-time.} of space-time corresponding to the two fixed points of $S^1/{\bf Z}_2$.
(In other words, in the eleventh dimension the first $E_8$ is localized on one of the fixed points, and the second $E_8$ is localized on the other fixed point.) Thus, the gauge couplings (in the first approximation) depend only on $R$ but not on ${\widetilde R}$. However, unless the two $E_8$'s are broken in the same way (that is, unless the instanton numbers for both $E_8$'s are the same), there is an upper bound on the size ${\widetilde R}$ of the eleventh dimension \cite{witten}. More precisely, the long wavelength expansion in eleven dimensions (that is, the effective field theory approach) breaks down unless ${\widetilde R}{\ \lower-1.2pt\vbox{\hbox{\rlap{$<$}\lower5pt\vbox{\hbox{$\sim$}}}}\ } R^4/\kappa^{2/3}$, where
$\kappa$ is the eleven dimensional gravitational coupling. (This holds for generic compactifications with unequal instanton numbers for two $E_8$'s.) Then one obtains a lower bound for the unification scale (for a fixed value of $M_P$):         
\begin{equation}
 M_{\small{GUT}}{\ \lower-1.2pt\vbox{\hbox{\rlap{$>$}\lower5pt\vbox{\hbox{$\sim$}}}}\ }
 \alpha M_P~.
\end{equation}
Note that this bound is the same\footnote{Intuitively one can understand this from the fact that upon compactification below ten dimensions the $Spin(32)/{\bf Z}_2$ and $E_8\otimes E_8$ heterotic vacua are related via T-duality.} as in the Type I theory with D9-branes, and is generically too high. On the other hand, we can consider compactifications with identical instanton numbers in both $E_8$'s to avoid such a bound. However, in this case symmetry breaking in the two $E_8$'s is the same, so we have two (largely) separated mirror worlds. {\em A priori} there is nothing inconsistent with such a picture, but the second $E_8$, which typically is thought of as the source of non-perturbative dynamics responsible for supersymmetry breaking and/or dilaton stabilization, is ``useless'' in this case. This poses the question of how to achieve other desired features in such a scenario\footnote{Here we should point out that {\em a priori} M5-branes (located in the bulk between the two worlds) can 
conceivably provide such non-perturbative dynamics. It would be interesting to understand this possibility better.}. 

{}There is, however, an alternative way of thinking about the above M-theory picture. Once we compactify enough dimensions, we can view such M-theory compactifications as (T-duals of) Type IIA orientifolds which are in turn dual to Type I$^\prime$ compactifications (where some gauge symmetries are non-perturbative). This leads us back to the brane world picture. Here we
would like to stress that starting from the brane world picture we might end up considering regimes of the theory (such as the strong coupling limit) where it is more advantageous to use some other, say, M-theoretic description. However, it is possible that one may have to use more     
than one description depending on a particular question one would like to address. With this in mind, we can take the brane world picture as a starting point to study various phenomenological aspects of the theory.

\section{Supersymmetry Breaking}

{}In this section we discuss various possible scenarios where the resulting low energy spectrum of string theory is non-supersymmetric. Here one can imagine (at least) three different scenarios.\\ 
$\bullet$ ({\em i}) Supersymmetry is broken at the tree level both on the brane and in the bulk. (Here, as well as in item ({\em ii}) below, we can {\em a priori} consider both tachyon free and tachyonic string vacua.)\\
$\bullet$ ({\em ii}) Classically supersymmetry is broken in the bulk but is unbroken on the brane - we refer to this scenario as ``Brane Supersymmetry''.\\ 
$\bullet$ ({\em iii}) Supersymmetry is broken dynamically on the brane ({\em e.g.}, via gaugino condensation).\\
In the following subsections we will consider these scenarios with the emphasis on certain points which will become relevant in the subsequent discussions. In particular, we will mostly be interested in properties of string perturbation expansion which can be understood in the context of some number of parallel D$p$-branes on which open strings can end.
Since we will ultimately be interested in having six compact spatial 
dimensions, we can consider Type I (or Type I$^\prime$) compactifications. In such a setup, gravity (described by the closed string sector of the theory) lives in the ten dimensional bulk, whereas we have some gauge theory (which is the low energy effective theory of the open string sector) living in the world-volume of the D$p$-branes.

{}Suppose we have a perturbative compactification with some number of unbroken space-time supersymmetries. Such a compactification is consistent to all orders in perturbation theory subject to string consistency requirements including tadpole cancellation conditions for various massless modes. These consistency constraints can be formulated at the one-loop level, and multi-loop consistency is then guaranteed as supersymmetry constrains possible higher loop corrections to various amplitudes in string theory. However, if we consider a background with completely broken space-time supersymmetry already at the tree level, additional complications arise at higher loop orders. In fact, in the following we discuss some of the model independent implications of such multi-loop properties of string perturbation expansion.
    
\subsection{Tree Level Supersymmetry Breaking}

{}Let us consider a generic perturbative compactification which is non-supersymmetric at the tree level. Let us first discuss tachyon free backgrounds which satisfy all one-loop consistency conditions (such as massless tadpole cancellation). In fact, let us confine our attention to the oriented closed string sector of the theory\footnote{The following discussion is effectively carried out for perturbative Type IIA/B or heterotic
compactifications. The conclusions, however, generically are also applicable in the context of Type I (or Type I$^\prime$) compactifications.}. The string perturbation expansion corresponds to summing over oriented genus $g\geq 0$ Riemann surfaces which for $g>1$ correspond to closed string loops. In particular, the cosmological constant is given by\footnote{The tree-level four dimensional cosmological constant vanishes 
as our starting point is string theory on $M_4\times X_6$, where $M_4$ is the four dimensional Minkowski space-time, and $X_6$ is some compact six dimensional manifold. Also, here, as well as in Appendix A, for the sake of simplicity we work in the string frame. It is not difficult to see, however, that our conclusions automatically apply to the Einstein frame as well.}
\begin{equation}
 \Lambda =\sum_{g=1}^\infty g^{2g-2}_s \Lambda_g~,
\end{equation}
where $\Lambda_g$ corresponds to the genus $g$ contribution. Note that $\Lambda_g$ is independent of $g_s$, and all the string coupling dependence is in the weight factor $g^{2g-2}_s$. 

{}Generically, the one-loop cosmological constant $\Lambda_1$ is non-zero, and it might depend on various ``parameters'' of the compactification such as the tree-level values of the radii of compact directions. However, these ``parameters'' in string theory are nothing but VEVs of the corresponding massless scalar fields in the low energy effective field theory. This implies that $\Lambda_1$ gives the scalar potential for such scalar fields (whose VEVs are arbitrary\footnote{Subject to certain requirements such as absence of tachyons.} at the tree level). Even if $\Lambda_1$ has a local minimum for the VEVs of the scalar fields that it depends on, $\Lambda_1$ is independent of at least one scalar field, namely, the dilaton VEV. This follows from the fact that the string coupling itself is given by the dilaton VEV $\phi$ via $g_s=e^\phi$. Thus, even if $\Lambda_1$ is non-zero, dilaton has a flat potential at the one-loop level as 
the corresponding weight $g^{2g-2}_s=1$ at genus $g=1$.

{}Next, consider the two-loop order. The two-loop cosmological constant depends on the dilaton VEV via $g^{2g-2}_s=g^2_s=e^{2\phi}$. However, if 
$\Lambda_2 > 0$, this dependence immediately leads to an inconsistency: the minimum of the scalar potential is at $\phi=-\infty$ which corresponds to the non-interacting string limit $g_s=0$. For any finite value of $g_s$ there is a dilaton tadpole, and the perturbative expansion breaks down. In fact, it is clear that perturbatively this ``runaway'' dilaton behavior is unavoidable unless $\Lambda_{g>1}\equiv 0$
(which generically implies that $\Lambda_1$ is also zero). This is the case in space-time supersymmetric compactifications. However, generically\footnote{Some non-supersymmetric
string backgrounds were recently argued to have zero cosmological constant to all loop orders in perturbation theory \cite{KKS}. It is not completely clear how precise these arguments can be made at higher genera. However, even if the cosmological constant in these theories vanishes to all loop orders, dilaton is still not stabilized perturbatively but is a modulus field.} non-supersymmetric backgrounds have non-vanishing cosmological constant at some finite loop order, so that the perturbation expansion breaks down. The key point
for our subsequent discussions is that even though supersymmetry is broken 
already (or absent) at the tree-level, dilaton is {\em not} stabilized 
perturbatively but rather has a runaway behavior toward the infinitely weak 
coupling limit.  

{}Here we can ask whether the divergences appearing at higher loop orders 
due to the dilaton tadpoles can somehow be regularized so that the 
perturbation theory is still meaningful. If so, at least naively, we could 
still hope to stabilize dilaton perturbatively. However, the following 
arguments indicate that such perturbative stabilization does not seem possible.
First, as pointed out in \cite{DiSe}, we do not expect the dilaton to be 
stabilized at weak coupling (so that perturbation theory is well defined) 
unless there is a large (or small) number in a given classical string 
background. Since there are no dimensionless couplings in string theory
(as all such couplings correspond to VEVs of various scalar fields), 
the appearance of such a large number is not generic.
One way around this could be, for instance, to have a (relatively) 
large number of D-branes. Then one could in principle imagine that the 
dilaton stabilization might occur at a weak string coupling value. 
However, the corresponding string theory would still not be weakly coupled. 
The point is that the effective (large $N$) 't Hooft coupling in the open 
string sector would still be of order one. On the other hand, even if we 
allow the possibility of a large (or small) number to be present in string 
theory, it still appears that the dilaton cannot be stabilized 
perturbatively. In appendix A, we give a formal argument indicating that,
 at any given loop order $g>2$, there are only three possibilities: 
({\em i}) the dilaton potential monotonously decreases as $\Sigma = g_s^{-2}$ 
increases (that is, we have runaway behavior); 
({\em ii}) there is a local maximum (but no minimum) of the dilaton potential;
({\em iii}) or $\exists \Sigma_g$ such that for $\Sigma>\Sigma_g$ we have a 
runaway potential, whereas for $\Sigma<\Sigma_g$ perturbation theory breaks 
down (and the genus $g$ cosmological constant
diverges) due to the fact that dilaton becomes tachyonic at genus $g-1$.  

{}Before we end this subsection, let us discuss non-supersymmetric backgrounds with tree-level physical tachyons\footnote{It is not difficult to construct such theories from perturbative heterotic string compactifications. In the case of Type I compactifications there are additional subtleties arising from the fact that here one needs to cancel both massless and tachyonic tadpoles which makes it non-trivial to find the corresponding solutions. Nonetheless, examples of anomaly free Type I (or Type I$^\prime$) compactifications with tree-level physical tachyons (in the closed string sector) can be constructed. Such theories were discussed in \cite{tachyon} in the context of compactifications on non-compact orbifolds ${\bf C}^3/\Gamma$. It is also straightforward to consider compact orbifolds $T^6/\Gamma$ along the lines of \cite{tachyon}. In fact, the massless spectra for the corresponding compact theories can be easily read off from the massless spectra in the non-compact cases given in \cite{tachyon}.}. 
The presence of tachyons indicates vacuum instability. In fact, the one-loop
cosmological constant is infinite due to tachyonic IR divergences if computed at the tachyonic point. Thus, for the perturbation expansion to make sense, it is necessary to find a stable vacuum at the tree level first. However, such a stable vacuum does not exist as the dilaton will have a runaway behavior already at the tree level. Thus, for simplicity let us consider a system with a single tachyonic field $\Phi$ coupled to the dilaton. A typical tree-level scalar potential reads:
\begin{equation}
 V_{\small{tree}} =-m^2 \Phi^2 + g_s^2 \Phi^4 +\dots~,
\end{equation}
where $m\sim M_s$ ($M_s=1/\sqrt{\alpha^\prime}$ is the string scale), and the ellipses stand for higher dimensional operators suppressed by powers of $g_s/M_s$. It is then clear that at the minimum
$|\Phi| \sim m/g_s$, and $-V\sim m^4/g_s^2$. Thus, once again we have runaway behavior for the dilaton $\phi$ in the direction of weak coupling, but unlike the tachyon free case we
have $V\rightarrow -\infty$ in this limit. 

{}As we see, perturbation expansion is not well defined in the above cases, one of the key reasons being that dilaton is not stabilized. Thus, for a string theory with supersymmetry broken already at the tree level to make sense perturbatively, it is necessary (albeit, perhaps, not sufficient) that dilaton is stabilized via some {\em non-perturbative} dynamics. Moreover, dilaton stabilization must occur at a scale which is at least somewhat larger than the supersymmetry breaking scale or else we will run into the same problem as before. In other words, generically dilaton stabilization must occur via some supersymmetry preserving dynamics such that the mass acquired by dilaton is larger than the supersymmetry breaking scale. In the language of effective field theory we can then integrate the dilaton field out and consider supersymmetry breaking (at lower energies) at a fixed coupling $g_s$. Then the loop expansion may have a chance to be well defined (albeit there are other subtleties at higher loop orders 
which we will not discuss here) in a given background.

\subsection{Brane Supersymmetry}

{}Next, we can imagine that supersymmetry is broken in the bulk at the tree level but classically supersymmetry is preserved on the brane. More precisely, the brane may feel supersymmetry breaking at some high mass levels but the massless spectrum at the tree level is supersymmetric. This scenario, which we refer to as brane supersymmetry, is possible in string theory, and we will discuss a simple example of such a string model in a moment. Here we would like to stress that, just as in the previous subsection, for the perturbation expansion to make sense we would still have to assume dilaton stabilization via some non-perturbative dynamics
(or else the closed string sector would not be well defined at higher loops). 

{}An example with brane supersymmetry can be constructed relatively easily in string theory \cite{Anto}. Thus, consider Type I$^\prime$ compactification on $(S^1)^6$ with D$p$-branes ($p<9$). Now consider supersymmetry breaking via the Scherk-Schwarz mechanism \cite{SS}.  
In the string language this corresponds to a freely acting orbifold which amounts to a half shift 
in the winding lattice corresponding to one of the circles (whose radius we will denote by $R$) accompanied by $(-1)^{F_L+F_R}$ in the closed string sector. (Here $(-1)^{F_L+F_R}$ acts as $+1$ in the space-time bosonic sectors (that is, NS-NS and R-R sectors), and as $-1$ in the 
space-time fermionic sectors (that is, R-NS and NS-R sectors)). This action breaks all supersymmetries in the closed string sector. More precisely, there are no massless closed string fermions, and the lightest states with the corresponding quantum numbers now have masses or order $1/R$ as they correspond to the Kaluza-Klein (momentum) modes on the circle. Let us now consider the open string sector. Let the above orbifold act on a circle {\em transverse} to the D$p$-branes. The open string modes in this direction are the winding modes whose masses are $\sim M^2_s R$. The action of the orbifold does not affect the massless spectrum (which remains supersymmetric) but modifies the massive winding spectrum (where at certain levels supersymmetry is broken). If $R\gg 1/M_s$, then these modes are very heavy and the effect of the orbifold on the supersymmetry in the massless spectrum is suppressed. Note that this is possible because the Scherk-Schwarz breaking orbifold acts in a direction transverse to the D-branes.

{}Had we orbifolded in a compact direction within the world-volume of D-branes, even at the massless level supersymmetry would be broken (for now the corresponding Kaluza-Klein momenta would feel supersymmetry breaking) with the boson-fermion splitting $\sim 1/R$. This latter case is an example of the scenario discussed in the previous subsection. 

{}Since classically supersymmetry (in the massless sector) is unbroken on the brane but is broken in the bulk, we expect quantum effects to trigger supersymmetry breaking on the brane.
For the particular example of brane supersymmetry discussed above, this issue was studied in \cite{Anto}. Field theoretic considerations are sufficient in this case to estimate supersymmetry breaking scale induced on the brane. Since it is ``gravity mediated'', we expect the corresponding scale to be suppressed by the Planck scale $M_P$, so that soft masses are of order $1/R^2 M_P$. Thus, if $R\gg M_P^{-1}$ we can have supersymmetry breaking scale on the brane which is much lower than the gravitino mass ($\sim 1/R$). However, this particular model is not generic. Thus, there are no chiral fermions in this model, and to obtain the latter we need to consider more involved compactifications. In fact, to obtain chiral fermions 
in four dimensions
generically one would have to consider compactifications on spaces of the Voisin-Borcea orbifold type, and then further orbifold by the action corresponding to Scherk-Schwarz breaking\footnote{The reason why some sort of orbifold is required here is that the action of the orbifold responsible 
for supersymmetry breaking must be a symmetry of the compact Calabi-Yau three-fold. To have a perturbative Type I (or Type I$^\prime$) compactification it is then required that this manifold is (a variation of) a Voisin-Borcea orbifold.}. In this framework one encounters additional branes whose world-volumes contain the circle which is orbifolded, and the gauge theory on these branes is non-supersymmetric. Supersymmetry breaking can then be ``gauge mediated'' to the original branes with tree-level supersymmetry, and the supersymmetry breaking scale on these original branes turns out to be of order $1/R$ (suppressed by the corresponding powers of the gauge coupling). So then $1/R$ cannot be much larger (to explain the hierarchy and naturalness) or much smaller (to be in agreement with experiment) than $1~{\mbox{TeV}}$.

{}Here we would like to point out another way of constructing theories with
tree-level brane supersymmetry. 
Thus, consider a ${\bf Z}_2\otimes {\bf Z}_2$ orbifold 
generated by two twists $\theta_L$ and $\theta_R$. Let the twists (acting as reflections on the corresponding compact coordinates) $\theta_L$ and $\theta_R$ be left-right asymmetric but
such that under the world-sheet parity reversal $\theta_L$ and $\theta_R$ are interchanged.
Then the twist $\theta_L\theta_R$ is left-right symmetric. (Note that $\theta_L$ and $\theta_R$ may be accompanied by shifts which are not going to be important in the following discussion except that they have to satisfy the same left-right symmetry properties as the corresponding twists.) Also note that the spectrum of Type IIB compactified on such an orbifold is left-right symmetric. Now consider the above orbifold 
with the action of $\theta_L$ and $\theta_R$ such that the corresponding Type IIB compactification is non-supersymmetric yet the twist $\theta_L\theta_R$ by itself preserves some number of space-time supersymmetries. Examples of such orbifolds have been discussed recently in \cite{KKS}. We can then construct an orientifold of this theory. The corresponding closed string sector is non-supersymmetric. However, the open string sector only ``feels'' the action of left-right symmetric twists. In this case the only such twist is $\theta_L\theta_R$ which by itself leaves some supersymmetries unbroken. Thus, the open string sector 
is (${\cal N} = 2$) supersymmetric in such a compactification. Hence we have 
tree-level brane supersymmetry (at all mass levels).
    
{}Before we end this subsection, let us make the following comment. Generically the cosmological constant in such Scherk-Schwarz type of models is of order $1/R^4$. Here we are not particularly concerned    
with the fact that it is large but rather would like to point out the runaway behavior for the VEV of the closed string scalar field which determines the radius $R$. In fact, in the limit $R\rightarrow \infty$ the model approaches a five dimensional {\em supersymmetric} theory (with vanishing cosmological constant). Such a runaway behavior (just as for the dilaton field) leads to inconsistencies at higher loops in perturbation theory. In fact, to have a consistent perturbation expansion it is necessary that the radius $R$ is stabilized at some finite value. It is not completely clear how to achieve this perturbatively. Just as for the dilaton field, it appears to be necessary that there is a {\em non-perturbative} potential generated for the radius modulus. 

\subsection{Dynamical Supersymmetry Breaking}

{}Finally, we can imagine a scenario where 
supersymmetry is dynamically broken on the brane. If we assume dilaton stabilization via some non-perturbative dynamics, then one of the simplest ways to obtain such a scenario is the old-fashioned gaugino condensate or some other mechanism for dynamical breaking of (global) supersymmetry in gauge theory. Such scenarios are not too much different from the standard heterotic picture albeit there are some new ingredients here which we would like to comment on briefly. 
Some detailed analyses of some of these possibilities can be found in \cite{peskin}.

{}Thus, let us first consider the ``gauge mediated'' scenarios. Here we can {\em a priori} have
the supersymmetry breaking dynamics taking place on the same set of branes as where the Standard Model fields reside. On the other hand, one can also imagine that this dynamics occurs on a different set of branes. Then supersymmetry breaking can be mediated by matter fields charged under gauge groups on both sets of branes, or by gauge fields (such as $U(1)$'s from the other branes) under which some of the Standard Model fields may be charged.

{}Similar possibilities occur for gaugino condensation type of mechanisms. Here we can have gravity mediated supersymmetry breaking within the same set of branes or communicated between two different sets. Here the following remark is in order. Suppose $M_s$ is not too different from the TeV scale (in which case we have the TeV string scenario \cite{TeV}). Then the gaugino condensate scale cannot be much lower than the string scale. On the other hand, since supersymmetry breaking is gravity mediated, it might appear that the corresponding scale will come out too low due to suppression by the Planck scale. However, to obtain the four dimensional $M_P$, we must assume that some of the compactified dimensions have radii much larger than inverse TeV (see section II). This implies that, besides the massless gravity modes, there are also a large number of the corresponding Kaluza-Klein modes which also participate in the mediation of supersymmetry breaking, so that the 
supersymmetry breaking scale can have the desired magnitude\footnote{We thank Gia Dvali for pointing this out to us.}. That is, instead of the ``old'' formula for the supersymmetry breaking scale in the observable sector, namely,     
\begin{equation}
 m_{\small{SUSY}} \sim {\Lambda^3/ M_P^2}~,
\end{equation}
where $M_P$ is the four dimensional Plank scale ($\sim 10^{19}~{\mbox{GeV}}$), and $\Lambda$ is the gaugino condensate scale, we must use a modified expression (which takes into account the contribution of the Kaluza-Klein modes) which reads
\begin{equation}
 m_{\small{SUSY}} \sim {\Lambda^3 / M^{\prime2}_P}~,
\end{equation}
where $M^\prime_P$ is the corresponding higher dimensional (``bulk'') Planck scale which roughly is related to the string scale as $M^\prime_P\sim M_s/g_s$ (up to a numerical coefficient
which depends on threshold corrections, {\em etc.}). For $\Lambda\sim M_s$ (which is around TeV) we can get the desired supersymmetry breaking scale.

\subsection{Signatures of SUSY at LHC}

{}We would like to end this section by briefly commenting on possible implications of the above scenarios for LHC. One can summarize this issue by saying that one way or another LHC should see some signatures of supersymmetry regardless of a particular scenario albeit details of such signatures do depend on a given case. For example, consider the case where supersymmetry is broken both in the bulk and on the brane at the tree level. Then we can imagine at least two possibilities: supersymmetry breaking occurs at the string scale $M_s$, or it occurs at some lower scale (such as the Scherk-Schwarz scale $1/R$). Generically, regardless of a particular scenario\footnote{Signatures of supersymmetry breaking via the Scherk-Schwarz mechanism have been studied in \cite{quiros}.}, the supersymmetry breaking scale cannot be much larger than TeV (or else it would be problematic to explain naturalness and hierarchy). So some signatures of superpartners should be detectable. In the former case $M_s$ itself cannot be much
 larger than TeV, so we have the TeV string scenario in this case. Similar conclusions hold for the 
brane supersymmetry scenarios as well as scenarios with dynamically broken 
supersymmetry (albeit certain details may be different). 

\section{Dilaton Stabilization}

{}In string theory gauge and gravitational couplings of the effective field theory are determined by VEVs of certain moduli. These are the scalar fields controlling the size (or the K{\"a}hler structure) of the compactified dimensions as well as the dilaton field. In this section we will concentrate on possible mechanisms for stabilizing the VEV of the dilaton field, that is, for generating non-trivial potential for $\phi$.

{}In space-time supersymmetric string vacua, dilaton is a modulus field to all orders in perturbation theory so its VEV is undetermined perturbatively. If we consider string vacua with supersymmetry broken already at the tree level, dilaton is not stabilized perturbatively. Following the discussion in section III,
we conclude that the dynamics responsible for dilaton stabilization must be 
intrinsically non-perturbative.

\subsection{Dynamical Mechanisms}

{}All known mechanisms of dilaton stabilization are related to some non-perturbative dynamics in the gauge sector of effective field theory\footnote{In M-theory there {\em a priori} exists a possibility that stabilization of moduli (including dilaton) may occur upon compactification on asymmetric orbifolds with frozen K{\"a}hler and complex structures\cite{ds}. In such a scenario moduli VEVs would be fixed at the eleven dimensional Planck scale. This would correspond to a strongly coupled string theory background in agreement with the conclusions of this section.}.    
All such mechanisms generically lead to dilaton stabilization at strong coupling values. In the following we will review the key reasons leading to this conclusion.

{}Instead of using the field $\phi$, for the following discussion it is convenient to work with a related field $S$ (which we will also refer to as the dilaton field) whose VEV determines the gauge coupling (at the string scale) in the sector of the theory responsible for dilaton stabilization: $\langle S \rangle = 1/\alpha + i\theta/2\pi$ (where $\theta$ is the corresponding vacuum angle). To begin with, let us consider dilaton stabilization in the context of ${\cal N}=1$ supersymmetric gauge theory. More precisely, let us first discuss scenarios where dilaton stabilization occurs via supersymmetry {\em preserving} dynamics. There are two known scenarios of this type.\\ 
$\bullet$ In the standard ``race track'' scenario \cite{kras}, non-perturbative superpotential (which is exponential in $S$) is generated via gaugino condensation in a pure super-Yang-Mills gauge theory. Dilaton stabilization then requires presence of at least two gauge groups giving rise to different exponentials in the superpotential. Thus, consider ${\cal N}=1$ supersymmetric gauge theory with $SU(N)\otimes SU(N^\prime)$ gauge group and no matter. The non-perturbative superpotential is given by:
\begin{equation}
 {\cal W}=a\exp(-2\pi S/N) + a^\prime \exp(-2\pi S/N^\prime)~,
\end{equation}  
where $a$ and $a^\prime$ are model dependent factors. For a suitable choice of $a,a^\prime$ the corresponding scalar potential has a supersymmetric 
minimum\footnote{Here, dilaton stabilization occurs without
breaking global supersymmetry. Moreover, the dilaton superfield is 
now massive (along with the SQCD fields which are all massive due to the 
mass gap). Once this system is coupled to supergravity,
there exists a locally supersymmetric vacuum in the vicinity of the globally 
supersymmetric one, as there is no Goldstino candidate required for 
local supersymmetry breaking.} at  
\begin{equation}
 {\mbox{Re}}(S)={1\over 2\pi }{N^\prime N\over N^\prime -N} \log\left|{aN^\prime\over a^\prime
 N}\right|~.
\end{equation}
Thus, generically, that is, if we assume $|N^\prime-N|\sim N$, the stabilized value of the gauge coupling is given by
\begin{equation}
 \alpha\sim 2\pi /N~.
\end{equation}
For this to be compatible with the gauge coupling at the unification scale (which is $\approx 1/24$) we must have $N{\ \lower-1.2pt\vbox{\hbox{\rlap{$>$}\lower5pt\vbox{\hbox{$\sim$}}}}\ }
10^2$, {\em i.e.}, very large gauge groups\footnote{Note that the effective 
't Hooft coupling 
is given by $\alpha N/2\pi\sim 1$, so that perturbative expansion is not 
valid. In particular, in the open string sector, the world-sheet expansion
in terms of Riemann surfaces with boundaries breaks down. We 
thank David Gross for pointing this out.} would be required for this mechanism to work\footnote{Such large gauge groups can {\em a priori} appear in non-perturbative string vacua. This possibility for dilaton stabilization was discussed in the context of F-theory in \cite{KL}.}. Here we should point out that {\em a priori} we could imagine a ``less generic'' scenario with, say, $N^\prime =N+1$ and relatively large $N$. Then the corresponding gauge coupling would be $\alpha\sim 2\pi\tau/N^2$, where $\tau\approx1/\log|a/a^\prime|$. Thus, if $\tau\sim 1$, we can obtain the desired value of the gauge coupling for $N\sim 10-15$. In the context of perturbative heterotic superstring this does not help as the total rank of the gauge group cannot exceed 22. However, in the context of non-perturbative sting theory such gauge groups are conceivable, so this mechanism might be viable if tuning $N$ close to $N^\prime$ can be achieved naturally\footnote{There is an additional objection to this mechanism, 
namely, that we have assumed $\tau\sim1$. Generically, however, for $N^\prime/N-1={\cal O}(1/N)$ (which we are assuming here) the coefficients $a$ and $a^\prime$ are almost identical (more precisely, $|a/a^\prime|-1={\cal O}(1/N)$) due to the almost universal couplings of the dilaton to the gaugino condensates through the gauge kinetic function. Nonetheless, there appears to be a way around this difficulty: we can start from two gauge groups with matter, and give masses to all of the matter fields via, say, {\em non-universal} Yukawa couplings (to an additional singlet field with a non-zero VEV). Then the corresponding coefficients $a,a^\prime$ can be such that $|a/a^\prime|-1={\cal O}(1)$.}. At present, however, no explicit string vacuum with such properties is known, and generically one does not expect such a mechanism to be easily realized.\\
$\bullet$ The second possibility \cite{gia} is to consider an ${\cal N}=1$ gauge theory with a simple gauge group, say, $SU(N)$ and charged matter. Then dilaton stabilization can occur due to an interplay between {\em quantum modification} of the moduli space and {\em tree-level} couplings of gauge invariants with additional gauge singlets. Let us briefly describe this mechanism\footnote{More details can be found in the original paper \cite{gia}. Also, our discussion here is in the context of global supersymmetry. The generalization (which does not change any of the conclusions) in the context of local supersymmetry can be found in \cite{klein}.}. Thus, consider a theory with $SU(N)$ gauge group and $N$ flavors $Q^i,{\tilde Q}_{\bar j}$ ($i,{\bar j}=1,\dots,N$). The gauge invariant degrees of freedom are mesons $M^i_{\bar j}\equiv Q^i {\tilde Q}_{\bar j}$, and baryons $B\equiv\epsilon_{{i_1}\dots {i_{N_c}}}Q^{i_1}\cdots Q^{i_{N_c}}$ and ${\tilde B}\equiv\epsilon^{{{\bar j}_1}\dots {{\bar j}_{N_c}}} 
{\tilde Q}_{{\bar j}_1}\cdots {\tilde Q}_{{\bar j}_{N_c}}$. The classical moduli space in this theory receives quantum corrections which can be accounted for via the following superpotential \cite{Seiberg}
\begin{equation}\label{non-pert}
 {\cal W}_{non-pert}=A\left(\det(M)-
 B{\tilde B}-\Lambda^{2N}\right)~,
\end{equation}
where $A$ is the Lagrange multiplier ($A\Lambda^{2N}=W_aW_a$ is the ``glue-ball'' field), and $\Lambda=\mu\exp(-\pi S/N)$ is the dynamically generated scale of the theory. Here $\mu\leq M_s$ is the scale below which all other degrees of freedom (such as heavy string modes and/or additional heavy fields which effectively decouple at energies below $\mu$) can be integrated
out. For simplicity, in the following we will set $\mu=1$.  
The quantum constraint follows from the F-flatness condition for the field $A$ and reads:
\begin{equation}\label{quantum}
 \det(M)-B{\tilde B}-\Lambda^{2N}=0~.
\end{equation}
Note that with just this constraint the dilaton is not stabilized. If, however, $\det(M)$, $B$ and ${\tilde B}$ are fixed via some other dynamics, then the quantum constraint (\ref{quantum}) will fix the dilaton VEV (provided that $0<\vert\det(M)-B{\tilde B}\vert<1$). In fact, the simplest possibility here is to consider a tree-level contribution to the superpotential (which can contain both renormalizable as well as non-renormalizable couplings). For instance, a generic tree-level superpotential (which respects all the symmetries of (\ref{non-pert})) is given by:
\begin{equation}
 {\cal W}_{tree}=Xf(\det(M),B{\tilde B})+Yg(\det(M),B{\tilde B})~,
\end{equation}
where $X,Y$ are singlet superfields both with $R$-charge 2, and $f,g$ are arbitrary polynomials of their arguments $\det(M)$ and $B{\tilde B}$. Note that the dilaton VEV is stabilized without breaking supersymmetry provided that the equation $f=g=0$ has isolated solutions with $0<\vert\det(M)-B{\tilde B}\vert<1$. The stabilized value of the gauge coupling is given by
\begin{equation}
 \alpha=2\pi \rho~,
\end{equation}     
where $\rho=-1/\log|\det(M)-B{\tilde B}|$. 
Generically, we expect $\rho\sim 1$ (albeit it is possible to lower $\rho$ via certain unnatural  and rather contrived adjustments of tree-level couplings - see \cite{gia}). Thus, dilaton stabilization occurs at strong coupling values $\alpha\sim 2\pi$.

{}Thus, both of the above mechanisms generically lead to dilaton stabilization in the strong coupling regime. We should point out that ``strong coupling'' here means ``coupling (that is, $\alpha$) of order one'', so that we (at least naively) do not expect a weakly coupled dual description. 
This might appear a bit troublesome as the observable effective field theory 
is weakly coupled at the electroweak scale. We will discuss a possible resolution (via the brane world picture) of this ``puzzle'' in a moment, but first we would like to make further comments on dilaton stabilization.

{}So far we have discussed the dilaton stabilization scenarios via supersymmetry preserving dynamics. It is reasonable to ask whether: ({\em i}) dynamical supersymmetry breaking and dilaton stabilization can possibly be due to the same non-perturbative dynamics; ({\em ii}) dilaton stabilization can be considered in the context of broken supersymmetry. Let us start with the second possibility. We have already pointed out that tree-level supersymmetry breaking does not seem to be acceptable in this context. What about dilaton stabilization in the context of {\em dynamically} broken supersymmetry, with the sources for dilaton stabilization and supersymmetry breaking being completely unrelated? This does not seem to be meaningful in the following sense. Dynamical supersymmetry breaking occurs via some dimensional transmutation which typically generates effects of the form $\exp(-c S)$ 
(where $c$ is a constant depending on the details of a particular scenario). This implies that, unless we can treat 
$S$ as a fixed coupling, the supersymmetry breaking scale 
(which is proportional to $\exp(-c S)$) will approach zero as $S\rightarrow\infty$
(that is, $\alpha\rightarrow 0$). Thus, we will have a runaway behavior for
$S$, and in the weak coupling limit supersymmetry will be restored. This
implies that we must ensure dilaton stabilization via some supersymmetry
preserving non-perturbative dynamics whose scale
$M_{\small{dilaton}}>>M_{\small{SUSY}}$. In this case supersymmetry
breaking is simply a small perturbation as far as the dilaton stabilizing
dynamics is concerned, so that we can (in the first approximation which is
good up to corrections ${\cal O}(M_{\small{SUSY}}/ M_{\small{dilaton}})$)
integrate the dilaton field out, and consider supersymmetry breaking at a
fixed value of the corresponding gauge 
coupling(s)\footnote{Here we would like to briefly comment on stabilization
of other moduli which is much less of a problem once the dilaton is
stabilized. Some of these other moduli may be stabilized together with the 
dilaton (as, say, in the supersymmetry preserving dilaton stabilization 
mechanism of \cite{gia}). The rest of them (or all of them) will 
generically be stabilized once supersymmetry is broken dynamically 
(as, say, in the usual supersymmetry breaking mechanism (with a fixed dilaton VEV) via 
gaugino condensation where all the geometric moduli acquire positive soft masses).}.

{}Let us now return to the question whether the same dynamics can be responsible for both supersymmetry breaking and dilaton stabilization. Note that here we are considering dynamical mechanisms where the strength of the corresponding non-perturbative effects is measured by  
the exponents of the form $\exp(-c S)$. From the previous discussions it should be clear that generically the stabilized value of the dilaton VEV is such that ${\mbox {Re}}(S)\sim 1/c$. This ultimately implies that $|\exp(-c S)|\sim 1$, and the supersymmetry breaking scale in this case should be $M_{\small{SUSY}}\sim M_s$. Here instead of $M_s$ we can have a different scale $\mu$ (below which the corresponding non-perturbative dynamics can be described in the language of effective field theory), but typically these two scales are not too different (where we assume that no fine tuning or any other unnatural adjustment is allowed). For example, in certain cases $\mu$ can be the anomalous $U(1)$ breaking scale which is somewhat (but not too much) smaller than $M_s$. Thus, generically we could {\em a priori} have such a mechanism in the context of the TeV string scenario. Here we should point out that realizing such a dynamical mechanism still appears to be rather non-trivial. One possibility here is to consider

gaugino condensation (where even a single gauge group without matter could suffice) in a theory with very {\em large} non-perturbative corrections to the K{\"a}hler potential \cite{BD} (also see, {\em e.g.}, \cite{MK}). It is then conceivable (although definitive quantitative conclusions are difficult to make) that such a mechanism can work. 
However, large corrections to the K{\"a}hler potential generically require the corresponding string theory to be strongly coupled. Thus, even in this case, we are led to the conclusion that dilaton stabilization (along with various considerations concerning supersymmetry breaking) generically requires the corresponding non-perturbative dynamics to occur in a sector of the theory which is strongly coupled at the string scale. This then implies that the string coupling $g_s$ itself cannot be small\footnote{This follows from the general relation $g^2_{\small{gauge}}\sim g^n_s /r^{p-3}$, 
where $n=1$ for D$p$-branes, and $n=2$ and $p=9$ in the heterotic string case. 
Here we assume that the dimensionless ``radius'' $r\geq 1$.}.   

\subsection{Why is the Effective Field Theory Weakly Coupled?}

{}The above remarks on dilaton stabilizations are troublesome in the context of  perturbative heterotic string theory. In particular, the effective field theory describing the Standard Model is weakly coupled (above the electroweak scale), and it is expected to be weakly coupled all the way to the GUT scale \cite{gut}. On the other hand, dilaton stabilization generically seems to require that the corresponding gauge theory be strongly coupled at the string scale. In perturbative heterotic string theory both the Standard Model and the ``hidden'' sector (supposedly responsible for dilaton stabilization, supersymmetry breaking, {\em etc.}) have identical (up to
integer coefficients of order one corresponding to possibly different current algebra levels) gauge couplings as they arise in the same (closed string) sector of the theory after breaking the original $E_8\otimes E_8$ or $SO(32)$ gauge symmetry. 

{}This is where the flexibility of the brane world picture becomes very useful. In this picture we can imagine that the Standard Model fields live in the word-volume of one set of $Dp$-branes with $p>3$ and the ``average'' dimensionless radius of the compact extra $p-3$ dimensions given by $r$. On the other hand, non-perturbative dynamics responsible for dilaton stabilizations may occur in the gauge theory coming from a different set of D$p_1$-branes with $p_1\geq 3$ and the ``average'' dimensionless radius of the compact extra $p_1-3$ dimensions given by $r_1$. (For $p_1=3$ we conventionally set $r_1=1$.) Then the (tree-level) gauge couplings (at the string scale) on these two sets of branes are given by:
\begin{eqnarray}
\alpha=g_s/2r^{p-3}~,~~~\alpha_1=g_s/2(r_1)^{p_1-3}~.
\end{eqnarray}  
Now if we would like to have $\alpha_1$ to be of order one (or even somewhat larger) yet $\alpha$ to be small (say, $\approx 1/24$), all we have to assume is that $g_s$ is in the strong coupling range, and the radii $r$ and $r_1$ are chosen appropriately. In particular, $r$ must be larger than $r_1$. Note that here it is {\em not} necessary to assume that $r\gg1$. More precisely, whether $r$ must be $\gg 1$ or not depends upon the value of $p$. For higher $p$-branes we could have $r$ only a few times bigger than 1. For example, if we take\footnote{Since $g_s$ is large and we are using tree-level relations, the following numerical estimates are meant to serve illustrative purposes only.} $\alpha_1=2\pi$ and $r_1 =1$, then we have $g_s=4\pi$, and $\alpha\sim1/24$ for $r\sim 3$ and $p=8$.   

{}Thus, the brane world scenario allows for the possibility of ``division of labor'', that is, different sets of branes can serve different purposes such as dilaton stabilization, supersymmetry breaking, giving rise to the Standard Model fields, and so on\footnote{Here we should point out that naively similar ``division of labor'' might seem to be possible 
in the old perturbative heterotic framework, where {\em a priori} different gauge 
subgroups could serve different purposes. However, as 
is well known \cite{BD}, the requirements of dilaton 
stabilization and weakness of the Standard Model 
gauge couplings do not seem to be compatible in 
this framework. This is because dilaton stabilization 
(as we discussed in the previous subsection) requires 
the corresponding gauge group to be strongly coupled 
at the string scale. In the perturbative heterotic framework 
this implies that all the other gauge subgroups, including 
those in the Standard Model, must also be strongly coupled at 
the string scale. This generically would lead to predictions 
(for the low energy gauge couplings) contradicting the data.}. 
The corresponding string theory is strongly coupled (in fact, most likely it is in the ``intermediate strong coupling'' regime so that it does not have a weakly coupled dual description), but on some of the branes the corresponding gauge theories can be weakly coupled due to relatively large (compared with the string length) compactification radii in the directions within the world-volumes of these branes. Here we note that in the Type I (or Type I$^\prime$) context various branes required in this scenario can simply be D-branes. In the language of non-perturbative heterotic string theory the branes responsible for, say, dilaton stabilization can be NS5-branes (while the Standard Model fields can come from the ``perturbative'' part of the spectrum).

\subsection{Varying Dilaton}

{}We would like to end this section by commenting on a possibility of the dilaton VEV {\em varying} over the compactification manifold, and its plausible phenomenological implications.
Thus, in perturbative heterotic or Type I compactifications the dilaton VEV is constant everywhere in the compact directions. In non-perturbative string theory this need not be the
case, and the dilaton VEV can vary point-wise in the compactification manifold. F-theory 
compactifications \cite{vafa} provide examples of this type. Thus, consider a generic F-theory compactification on a Calabi-Yau $(n+1)$-fold ${\cal M}_{n+1}$ which is an elliptic $T^2$ fibration over a base ${\cal B}_n$. Four dimensional ${\cal N}=1$ supersymmetric compactifications are obtained upon compactifying on a Calabi-Yau four-fold (with $SU(4)$ holonomy). Let $z_i$ ($i=1,\dots,n$) be a set of complex coordinates parametrizing ${\cal B}_n$, and let $\tau(z)$ denote the modular parameter of the fiber $T^2$ as a function of $z$. By definition, F-theory compactified on ${\cal M}_{n+1}$ is Type IIB on ${\bf R}^{9-2n,1}\times {\cal B}_n$ with 
$\tau(z)= a(z) + i e^{-\phi(z)}$, where $a$ is the R-R scalar, and $\phi$ is the dilaton field. 
If we consider an elliptic fibration with the modular parameter $\tau$ varying over the base ${\cal B}_n$ then the dilaton VEV $\phi$ (and, therefore, the Type IIB string coupling constant $g_s$)  
also varies over the base. 

{}The fiber can degenerate in an $n-1$ complex dimensional subspace ${\cal C}_{n-1}$ of ${\cal B}_n$ so that $\tau(z)$ goes to $i\infty$ (that is, $g_s\rightarrow 0$). To cancel the corresponding anomalies we must include D7-branes whose world-volume fills ${\bf R}^{9-2n,1}\times {\cal C}_{n-1}$. The location of these D7-branes is then given by a point on ${\cal B}_n$ in the direction transverse to ${\cal C}_{n-1}$. Near this point we can use perturbative string description in terms of open strings ending on D7-branes. However, away from this point such description breaks down. Moreover, generically we can have mutually non-local D7-branes (corresponding to various degenerations of the fiber) which cannot be simultaneously described in perturbation theory. Also, for compactifications on Calabi-Yau four-folds we must include $\chi/24$ three-branes to cancel other anomalies \cite{SVW}, where $\chi$ is the Euler number of the four-fold.

{}Note that perturbative Type IIB orientifolds with D7-branes (as well as D3-branes) can be viewed as a special limit of F-theory compactifications where the modular parameter is constant (and such that the corresponding string coupling is small) almost everywhere on the base \cite{sen}. In fact, the ``exceptional'' points correspond to orientifold 7-planes. Thus, perturbative Type I$^\prime$ vacua with D7-branes (which correspond to perturbative 
Type IIB orientifolds with D7-branes) can be viewed as special limits of F-theory compactifications. More generic non-perturbative vacua therefore generically have the dilaton VEV varying over the compactification manifold.

{}Varying string coupling in the brane world picture may have interesting phenomenological implications a few of which we would like to briefly mention here (although a more detailed analyses would be required to determine their plausible experimental signatures).\\
$\bullet$ The gravitational coupling on the brane and in the bulk can be 
quite different. This might have implications for the bulk effects on the 
brane physics in cosmology ({\em e.g.}, big-bang nucleosynthesis) and 
astrophysics (especially in the cases where $M_s$ is as low as 
$\sim 1~{\mbox{TeV}}$).\\      
$\bullet$ The dilaton VEV can vary from one set of branes 
to another. It is then conceivable that different fields in the Standard Model couple to gravity with different strengths. This would be especially interesting in the TeV string scenario where gravitational effects can become important at energy scales around TeV (which might be accessible to LHC).\\
$\bullet$ One can also imagine that difference in various gauge couplings (say, for the Standard Model fields) can at least partially be due to dilaton varying between different sets of branes (and not only due to the difference in the compactification volume factors). This might have implications for understanding the gauge and gravitational coupling unification as well as in the context of various remarks on dilaton stabilization made in this section.

{}It would be interesting to understand plausible phenomenological implications of varying dilaton in a more quantitative fashion. This most likely would require further developing
model building tools for constructing four dimensional F-theory compactifications.

\section{Outlook}

{}In this section we would like to summarize various discussions in the previous sections. 
We have revisited some of the issues in string phenomenology, namely, we have 
discussed the implications of the following constraints (necessary for any string model to be phenomenologically viable).\\
$\bullet$ The gauge and gravitational couplings must unify at the string scale.\\ 
$\bullet$ The dilaton VEV must be stabilized. (In discussing this point we have considered various scenarios of supersymmetry breaking.)\\
$\bullet$ The low energy effective field theory (corresponding to the Standard Model) 
is weakly coupled below some scale.\\
These constraints generically seem to imply that in the context of string theory 
the brane world scenario is favored as a description of nature. Within this scenario our four dimensional world resides inside of a $p$-brane (or a wall-like object reminiscent of a $p$-brane). More precisely, the Standard Model fields are localized on such a $p$-brane (or a set of overlapping branes), whereas gravity propagates in a larger (10 or 11) dimensional bulk.
Moreover, the corresponding string theory appears to be strongly coupled. More precisely, the string coupling $g_s{\ \lower-1.2pt\vbox{\hbox{\rlap{$>$}\lower5pt\vbox{\hbox{$\sim$}}}}\ }1$, but the theory has no weakly coupled dual description. Weakly coupled effective field theory (corresponding to the sector describing the Standard Model fields) then arises due to somewhat large compactification scale in the directions within the corresponding brane(s).

{}For illustrative purposes, let us discuss a possible realization of such a brane world scenario in string theory. For concreteness we will do so within the context of Type I compactifications to four dimensions. However, we stress that other dual descriptions are just as valid. In fact, to gain as much insight 
as possible into how a specific brane world scenario can describe nature, 
it may be necessary to use different dual descriptions
(such as M-theory, F-theory, non-perturbative heterotic, {\em etc.}), each description
providing a valuable viewpoint of certain particular aspects of the physics.

{}Thus, let us consider a Type I compactification on a Calabi-Yau three-fold\footnote{For recent developments in four dimensional Type I compactifications see \cite{typeI,3gen,ST}.}. 
In this theory we have D9-branes, and perturbatively we can have three different sets of D5-branes\footnote{Examples of such 
compactifications are toroidal orbifolds with a ${\bf Z}_2 \otimes {\bf Z}_2$
subgroup. ``Semi-realistic'' examples (with three chiral families) of this type
have been constructed in \cite{3gen}.}. 
Let $X^0,\dots,X^3$ be the coordinates corresponding to the four dimensional Minkowski space-time ${\bf R}^{3,1}$, and $X^4,\dots,X^9$ be the compact coordinates. Let us denote the three possible sets of D5-branes as D$5_i$-branes with $i=1,2,3$, and let the corresponding world-volumes be filling (along with ${\bf R}^{3,1}$) the compact directions $X^{3+i},X^{4+i}$. 
Also, let $R_i$ be the ``sizes'' of the compact directions $X^{3+i},X^{4+i}$. 
The Standard Model fields can come from the D$5_1$-branes. This implies that $R_1$ should be relatively large  compared with $1/M_s$ (but {\em not} many orders of magnitude larger than $1/M_s$). This is required to obtain {\em weakly} coupled effective field theory corresponding to the Standard Model as we are going to have relatively large (${\ \lower-1.2pt\vbox{\hbox{\rlap{$>$}\lower5pt\vbox{\hbox{$\sim$}}}}\ } 1$) string coupling to allow for dilaton stabilization. The latter can occur via non-perturbative dynamics arising in the gauge theory living on, say, the D$5_2$-branes with $R_2\sim 1/M_s$ so that the corresponding effective field theory is {\em strongly} coupled at the string scale. 
The role of the D$5_3$- and D9-branes would depend on the values of $R_3$. If $R_3\gg 1/M_s$, then the corresponding gauge theories are much more weakly coupled than that on the D$5_1$-branes. Thus, from the viewpoint of the Standard Model, the D$5_3$- and D9-brane charges can serve as horizontal (or flavor) symmetry quantum numbers. Note that in this case supersymmetry breaking dynamics would have to occur on either the D$5_1$- or D$5_2$-branes, or at the tree level.
Alternatively, $R_3$ can be comparable to $R_1$ or $1/M_s$. In this case supersymmetry breaking can come from the dynamics on either of the 
D$5_3$- and D9-branes. 

{}We can summarize the key point of this discussion by saying that the brane world picture appears to be flexible enough to accommodate various (if not all) phenomenological requirements that used to be so difficult to fit together in, say, old perturbative heterotic framework.
Flexibility of the brane world scenario has its flip side as well, namely, it might lead to the lack (perhaps, to certain degree) of predictive power. This is especially pressing as the corresponding string theory appears to be in the strong coupling regime so that quantifying various qualitative statements can be rather difficult (even given the correct vacuum
corresponding to our universe). Perhaps, one should directly address certain 
fundamental questions, such as the cosmological constant problem, 
which at the end of the day would ultimately have to be answered anyway.

\acknowledgements

{}We would like to thank Philip Argyres, Nima Arkani-Hamed, Itzhak Bars, Savas Dimopoulos, Michael Dine, Gia Dvali, David Gross, Petr Ho{\u r}ava, Shamit Kachru, Pran Nath, 
Hirosi Ooguri, Michael Peskin, Joe Polchinski, John Schwarz, 
Stephen Shenker, Gary Shiu, Eva Silverstein, Cumrun Vafa and Piljin Yi for valuable discussions.
The research of S.-H.H.T. was partially supported by the
National Science Foundation. 
The work of Z.K. was supported in part by the grant NSF PHY-96-02074,
and the DOE 1994 OJI award.
Z.K. would also like to thank Albert and Ribena Yu for financial support.

\appendix

\section{Perturbative String Expansion}

{}In this appendix we give an argument that dilaton stabilization cannot occur perturbatively
in tachyon free string vacua with supersymmetry broken at the tree level. Thus, as in subsection
A of section III, let us consider such a vacuum in the context of closed string world sheet expansion. Let $\Sigma\equiv g_s^{-2}=e^{-2\phi}>0$. 
The cosmological constant is given by:
\begin{equation}
 \Lambda =\sum_g \Sigma^{1-g} \Lambda_g~,
\end{equation}
where $\Lambda_g$ are independent of $\Sigma$. 

{}As we have already pointed out in section III, at the two-loop order dilaton has a runaway
potential: $V_2=\Lambda_1 +\Sigma^{-1} \Lambda_2$. More precisely, if $\Lambda_2>0$ then
dilaton will roll to the infinitely weak coupling values: $\Sigma\rightarrow\infty$. Note that if $\Lambda_2<0$ then dilaton is tachyonic for all values of $\Sigma$ at this order, and the perturbation theory breaks down at the three-loop order. Thus, let us assume that $\Lambda_2>0$. In this case we have dilaton tadpole so that we expect divergences at higher loop orders. These divergences will occur from the world-sheet configurations where a 
$(g+2)$-genus Riemann surface factorizes into a genus $g$ surface connected to a genus 2
surface by a long thin tube. Then dilaton contribution to such a configuration is divergent due
to a two-loop tadpole. However, the lowest order at which such a divergence occurs is the four-loop order as at three loops the two-loop tadpole is multiplied by a one-loop tadpole, which is vanishing, corresponding to the genus one surface arising in the factorization limit. 

{}Thus, we do not have to worry about such divergences at three loops. Let 
us then 
assume that $\Lambda_3$ is well defined. Note that dilaton stabilization cannot occur at this
order: if $\Lambda_3>0$, then we still have runaway as $\Sigma\rightarrow \infty$; on the other hand, if $\Lambda_3<0$, then the extremum at $\Sigma =-2\Lambda_3/\Lambda_2$ corresponds to the maximum of the three-loop potential $V_3=\Lambda_1 +\Sigma^{-1}
\Lambda_2 +\Sigma^{-2} \Lambda_3$. Also note that if $\Lambda_3<0$ then for $\Sigma<\Sigma_*$ dilaton becomes tachyonic, where $\Sigma_*\equiv -3\Lambda_3/\Lambda_2$.

{}Next, let us consider the four-loop order. Let us assume that we can somehow regularize $\Lambda_4$ so that it is well defined. Then we have three possibilities: ({\em i}) if $\Lambda_3>0$ and $\Lambda_4>0$, then the dilaton potential $V_4=\Lambda_1 +\Sigma^{-1}
\Lambda_2 +\Sigma^{-2} \Lambda_3+\Sigma^{-3} \Lambda_4$ monotonously decreases as $\Sigma$ increases (that is, we have runaway behavior); ({\em ii}) if $\Lambda_4<0$ (regardless of whether $\Lambda_3$ is positive or negative), then
there is a local maximum (but no minimum) of the
dilaton potential; ({\em iii}) if $\Lambda_3<0$ and $\Lambda_4>0$, then naively we have a local minimum at $\Sigma_{\small{min}}=\Lambda^{-1}_2\left[-\Lambda_3-\sqrt{\Lambda_3^2-
3\Lambda_2\Lambda_4}\right]$ 
(provided that $\Lambda_4 <\Lambda_3^2/3\Lambda_2$). However, in the last case we run into an inconsistency. The point is that this minimum is located at
$\Sigma_{\small{min}}<\Sigma_*$. However, at the three-loop order dilaton becomes tachyonic for $\Sigma<\Sigma_*$, and the four-loop contribution $\Lambda_4$ diverges due to this tachyonic contribution. That is, perturbation theory breaks down at four loops in this case unless
$\Sigma>\Sigma_*$, but this corresponds to the runaway behavior.

{}It should be clear how to generalize the above argument to all genera. 
Thus, we can use an inductive argument where one assumes that there is no 
local  minimum at some genus $g(\geq 2)$, and then shows that at genus 
$g+1$ there is either only a runaway potential or a local maximum (plus 
the runaway behavior), but no local minimum. In particular, just as we saw 
at the four-loop order, at any genus $g>3$ in one of the possible cases 
naively there is a local minimum, but the latter is always
at a value where perturbation theory breaks down due to the fact that 
dilaton becomes tachyonic at genus $g-1$ for this value of $\Sigma$. 
In this argument one assumes that all
the divergences due to dilaton tadpoles can be regularized. 
This is not a restrictive assumption as perturbation theory would 
otherwise break down already for this reason alone.

\end{document}